\documentclass[sigconf]{acmart}

\usepackage{graphicx}
\usepackage{color}
\usepackage{amsmath}
\usepackage{amsmath,amssymb}
\usepackage{float}
\usepackage{mathtools, cuted}
\usepackage{bytefield}
\usepackage{subfig}
\usepackage{multirow}
\usepackage{siunitx}

\usepackage{algorithm,algpseudocode}
\algtext*{EndFor}
\algtext*{EndIf}
\algtext*{EndProcedure}

\AtBeginDocument{%
  \providecommand\BibTeX{{%
    \normalfont B\kern-0.5em{\scshape i\kern-0.25em b}\kern-0.8em\TeX}}}

\copyrightyear{2020} 
\acmYear{2020} 
\setcopyright{acmcopyright}
\acmConference[AutoSec '20]{Proceedings of the Second ACM Workshop on Automotive and Aerial Vehicle Security}{March 18, 2020}{New Orleans, LA, USA}
\acmBooktitle{Proceedings of the Second ACM Workshop on Automotive and Aerial Vehicle Security (AutoSec '20), March 18, 2020, New Orleans, LA, USA}
\acmPrice{15.00}
\acmDOI{10.1145/3375706.3380554}
\acmISBN{978-1-4503-7113-1/20/03}

\begin{document}
\fancyhead{}
\title{Secure Traffic Lights: Replay Attack Detection for Model-based Smart Traffic Controllers}

\author{Pratham Oza}
\authornote{Both authors contributed equally to this research.}
\affiliation{%
  \institution{Virginia Tech}
}
\email{prathamo@vt.edu}

\author{Mahsa Foruhandeh}
\authornotemark[1]
\affiliation{%
  \institution{Virginia Tech}
}
\email{mfhd@vt.edu}

\author{Ryan Gerdes}
\affiliation{%
  \institution{Virginia Tech}
}
\email{rgerdes@vt.edu}

\author{Thidapat Chantem}
\affiliation{%
  \institution{Virginia Tech}
}
\email{tchantem@vt.edu}

\renewcommand{\shortauthors}{}

\begin{abstract}
Rapid urbanization calls for smart traffic management solutions that incorporate sensors, distributed traffic controllers and V2X communication technologies to provide fine-grained traffic control to mitigate congestion. As in many other cyber-physical systems, smart traffic management systems typically lack security measures. This allows numerous opportunities for adversarial entities to craft attacks on the sensor networks, wireless data sharing and/or the distributed traffic controllers. We show that such vulnerabilities can be exploited to disrupt mobility in a large urban area and cause unsafe conditions for drivers and the pedestrians on the roads. Specifically, in this paper, we look into vulnerabilities in model-based traffic controllers and show that, even with state-of-the-art attack detectors in place, false-data injection can be used to hamper mobility. We demonstrate a replay attack by modeling an isolated intersection in VISSIM, a popular traffic simulator and also discuss countermeasures to thwart such attacks. 
\end{abstract}


\maketitle

\section{Introduction}
\label{sec::intro}


More people are moving into urban areas for better economic and social prospects~\cite{urbanization}. This is accompanied by increasing demands on urban transportation infrastructure which is typically at capacity and often cannot cope with this surge in demand. As a result, city-dwellers often face traffic congestion. On an average, commuters face about 42 hours of delays in a year due to traffic congestion~\cite{cookson2017inrix}. Such delays cause wasted time, negative impacts to the socioeconomic growth of the area, increased pollution and fuel expenses~\cite{fhwa}.

Smart cities provide a promising framework to tackle issues pertaining to urban resources~\cite{varga2017architecture}. Many traffic control strategies have been proposed that rely on sensors, on-road detectors, cameras and/or other road-side units (RSUs) to measure real-time traffic flow characteristics such as queue lengths, vehicle speeds, traffic density, etc~\cite{tonguz2019harnessing}. The measured data is then communicated to a traffic controller that adaptively manages the traffic flow as per the deployed control strategy~\cite{zammit2019realtimemodel, 8864584}.
However, most of these techniques are deployed on top of the legacy traffic infrastructure~\cite{day2019leveraging} which was built without security in mind. This makes the intelligent traffic management system, like many other cyber-physical systems, vulnerable to attacks~\cite{ghena2014green,kelarestaghi2019intelligent}. Adversarial entities can exploit the weaknesses of these systems using, for example, replay attacks (i.e., provide controllers with spurious sensing information~\cite{feng2018falsedata}) or disrupt measurements and communications (i.e., denial-of-service (DoS) attacks~\cite{li2016assessing}) to cause phantom traffic jams or otherwise degrade the quality of (transportation) service~\cite{liu2011arterial}.

Model-based traffic control approaches have been fairly popular in estimating traffic queues and controlling the traffic flow. Traditional approaches such as UTOPIA~\cite{mauro1990utopia}, PRODYN \cite{henry1984prodyn}, and RHODES \cite{rhodes}, etc., as well as newly developed MPC-based traffic control algorithms~\cite{lin2009simplified} rely on traffic models and state-space estimations to predict traffic flow and hence establish a finer control over the traffic. Such model-based controllers not only provide real-time traffic control from received information about vehicle density and speed but also provide self-adaptive control strategy that combines current measurements as well as future estimations of the traffic characteristics~\cite{zammit2019realtimemodel, fernandez2015modeldriven}. Since such systems still rely on measurements from insecure sensors, they need to accompanied with attack detectors that monitor the estimates made by the model. These detectors can detect malicious activities if the decisions do not corroborate the observations~\cite{pasqualetti2013attack}. Our key contributions are:
\begin{itemize}
 \item We propose a replay attack, which is a subset of false-data injection attacks, on the inductive loop detector used in transportation systems to estimate vehicle flow.
 \item We explore the shortcomings of the existing model-based attack detectors.
 \item We provide a threshold-based countermeasure and validate its performance with VISSIM, a real-world traffic simulator. 
\end{itemize}

In this work, we discuss the system model as a four-way traffic intersection and provide a state-space model in Sec.~\ref{sec::system_model}. Then, in Sec.~\ref{sec::threat_model} we design a replay attack on the inductive loop detectors. We demonstrate the shortcoming of model-based detectors in detecting our attack in Sec.~\ref{sec::cm:model}. We propose a threshold-based approach as a countermeasure in Sec.~\ref{sec::cm:thrsh} and then conclude in Sec.~\ref{sec::conc}. 



\section{System Model}
\label{sec::system_model}
This section delves into the traffic parameters and state-space models to mimic a four-way intersection traffic controller.

\subsection{Traffic, Intersection and Infrastructure}
Most traffic controllers in the U.S. deploy inductive loop detectors to collect traffic-related information~\cite{guerrero2018sensor}. Occupancy sensor readings obtained from these inductive loop detectors are then used to establish control over the traffic lights~\cite{burger2013considerations}. In this paper, we consider an isolated four-way intersection as shown in Fig.~\ref{fig::figure_1}a. Traffic lights are situated at the end of all four lanes downstream and are controlled using our model-based traffic controller. This traffic controller acquires real-time measurements from inductive loop detectors located upstream. For simplicity, we consider that the vehicles use the intersection only to go straight (no left/right turns). Adding turning lanes and vehicles simply requires extending the system model with multiple traffic lights. Incoming traffic within each lane has the following characteristics:
\begin{itemize}
    \item Intensity ($I_{t}$)[v/h]: Average rate of vehicles entering a given lane at a time instant $t$.
    \item Saturation Flow ($S$)[v/h]: The maximum number of vehicles per hour that can flow through the intersection depending on the lane capacity.
    \item Green Light Duration ($z$)[s]: The time for which the vehicles from a lane can access the intersection.
\end{itemize}


The traffic controller decides the green-yellow-red timing schedule for the traffic lights located at the intersection. The controller follows a \textit{cycle time ($T_c$[s])} that defines the time window within which each traffic light turns green at least once. At the beginning of each $T_c$, the controller decides the schedule for the next cycle window as per the control algorithm and the process repeats at the end of the cycle. 

A vehicle entering the lane activates the inductive loop installed on the road that generates a signal variation whose duration depends on how long a vehicle stays over the detector. The amount of time the detector remains activated within a given sample period determines the \textit{occupancy rate $(O[\%])$} of the sensor. When a vehicle passes over the detector at high speed, the occupancy rate reported would be a lower value since the vehicles spend less time over the detector. This value increases as the speed of the passing vehicle decreases, and eventually saturates at 100\% when the vehicle is stationary on top of the detector. 

While multiple loop detectors may be located on the road to measure different traffic characteristics, in practice, only the one located farthest from the intersection is used to measure the flow of the traffic as it has the least probability of saturation~\cite{fhwa_loop}. Therefore, our traffic model considers the loop detectors situated at the farthest end of our incoming lanes, as shown in Fig.~\ref{fig::figure_1}a for traffic estimation. Our model is communication-agnostic and the measurements may be reported using Dedicated Short Range Communication (DSRC), CV2X, etc., while assuming that the communication channel is secure. We now provide a state-space model that closely represents our adaptive traffic control system.


\subsection{Model-based Self-Adaptive Traffic Controller}

Among the various application-specific models that have been developed to analyze traffic characteristics, the model presented in~\cite{dunik2006state} provides a state-space model to estimate the parameters of the traffic model based off inductive loop measurements~\cite{zammit2019realtimemodel}. We use the model in~\cite{dunik2006state} to adjust the green light duration at the intersection. 


Traffic control within each lanes of an intersection can be characterized by estimating the development of vehicle queues in each lane over time. By utilizing the occupancy rate measurements provided by the sensors (also known as the measurable quantities) the states of the system ($X_t$) at a given time instant $t$ need to be formulated to describe how the queues evolve with the measurements and the control decisions for the green light. The state-space equations are defined as
\begin{equation}
\label{eq::x}
    X_{t} = \left[Q_{t}\,\, ,O_{t}\right]^{T},
\end{equation}
where $Q_{t}$ is the current estimated queue length in a given lane and $O_{t}$ is the occupancy of the inductive loop detector at time $t$.

Similarly, $y_{t}$ defines the observable quantities in the system that influence the evolution of the queues in all lanes and is given by
\begin{equation}
\label{eq::y}
    y_{t} = \left[Y_{t}\,\, ,O_{t}\right]^{T},
\end{equation}
where $Y_{t}$ is the number of vehicles being dispatched within a given time window of $T_c$ due to the controlled green light.

The queue length within each lane also depends on the saturation flow of the lane. Saturation flow ($S$ [veh/h]), as explained earlier, provides the maximum number of vehicles that can enter into the lane due to the physical capacity of the lanes and intersection. Hence, to relate the queue development with the saturation flow, a parameter $\delta$ is introduced. $\delta$ is set to zero when the incoming vehicle flow is less than the saturation flow and is set to unity, otherwise. The value of $\delta$ is updated at the start of each traffic cycle window and ensures that the incoming flow never exceeds the saturation flow.

\begin{figure*}
  \centering
\begin{tabular}{ccc}
\multirow{2}{*}{\includegraphics[width=0.33\textwidth]{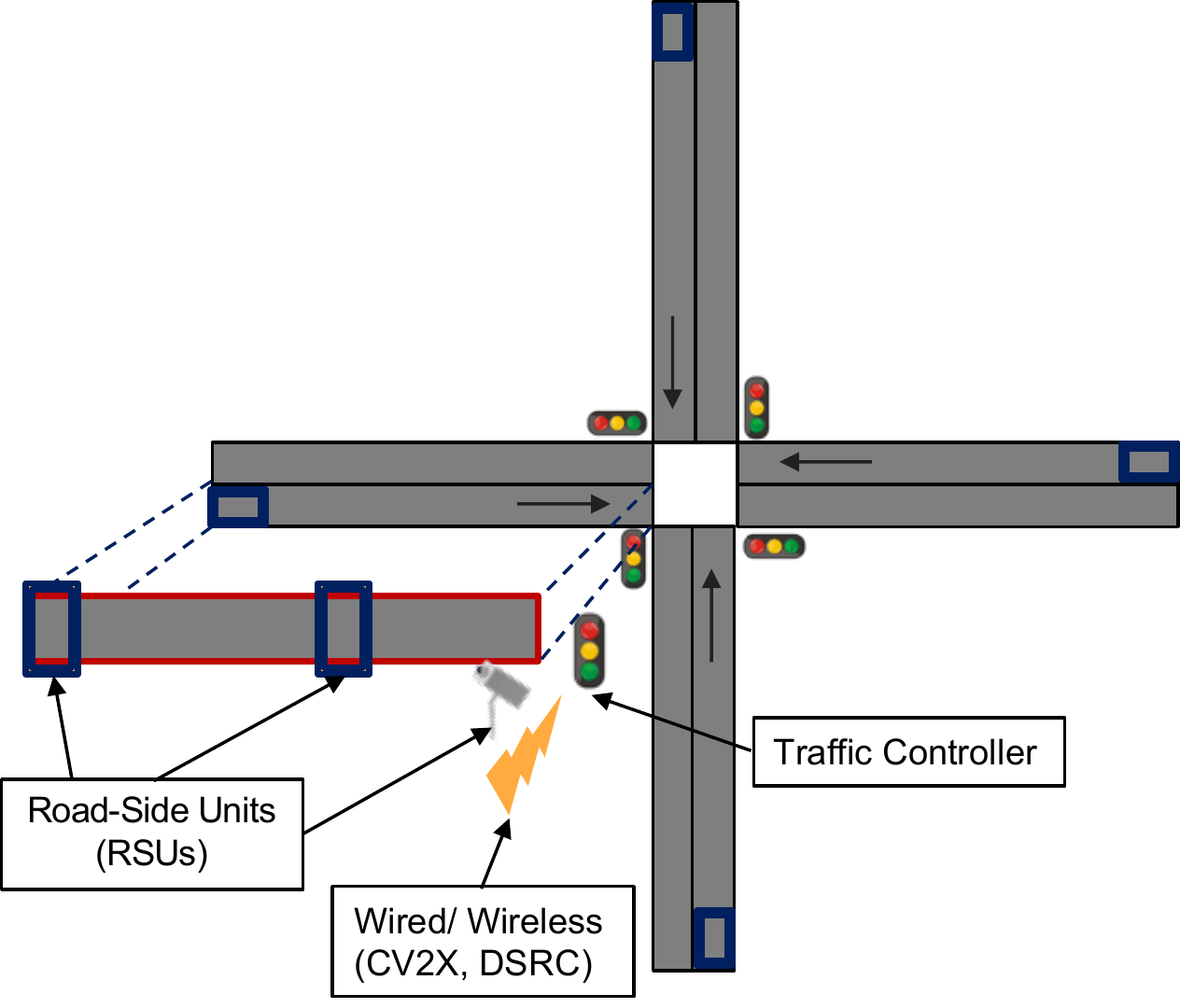}}&
\includegraphics[width=0.33\textwidth]{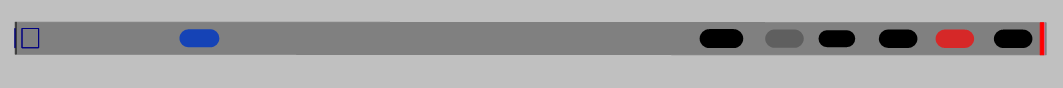}&
\multirow{2}{*}{\includegraphics[width=0.33\textwidth]{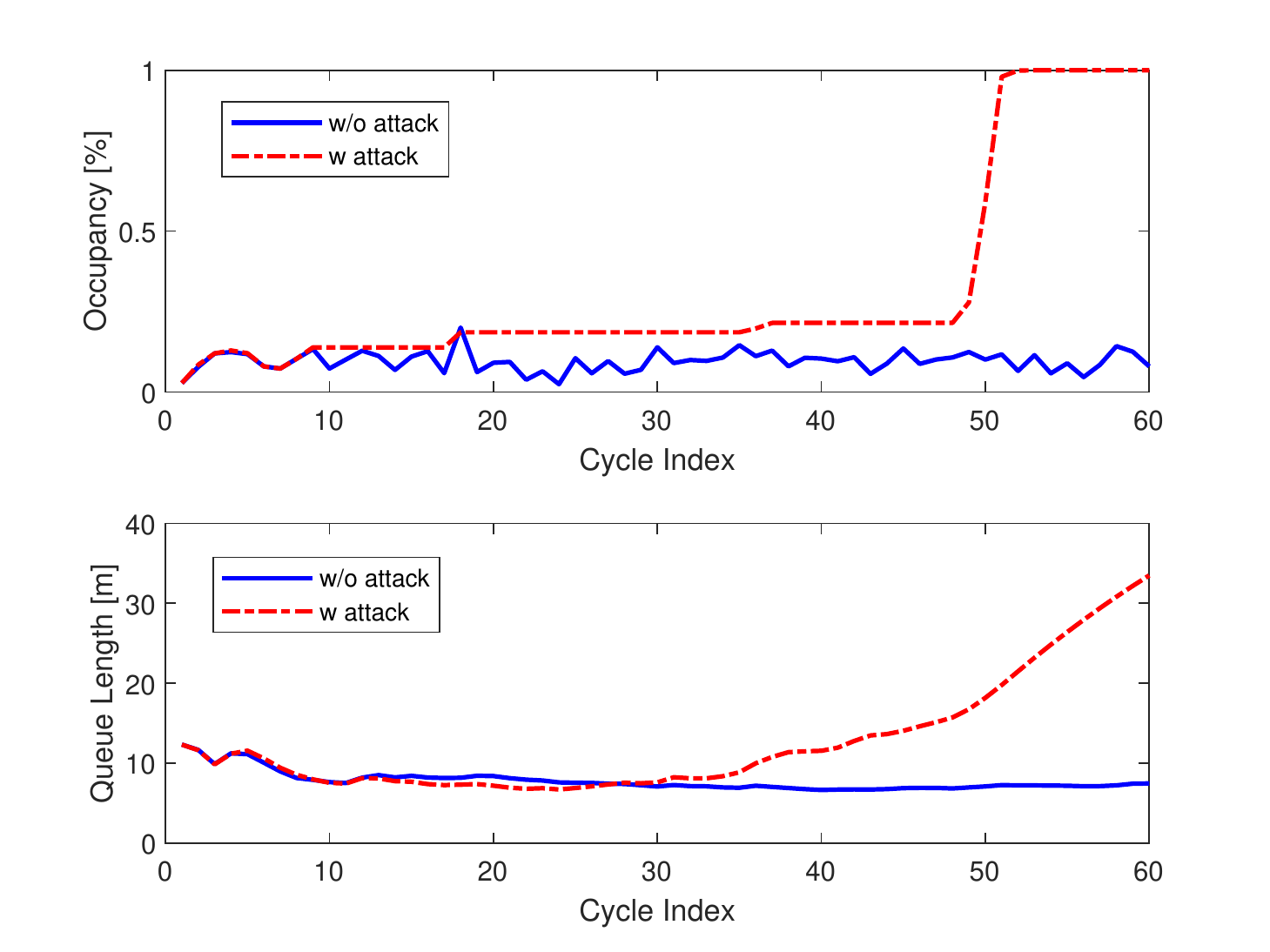}}\\
&(b)&\\
&\includegraphics[width=0.33\textwidth]{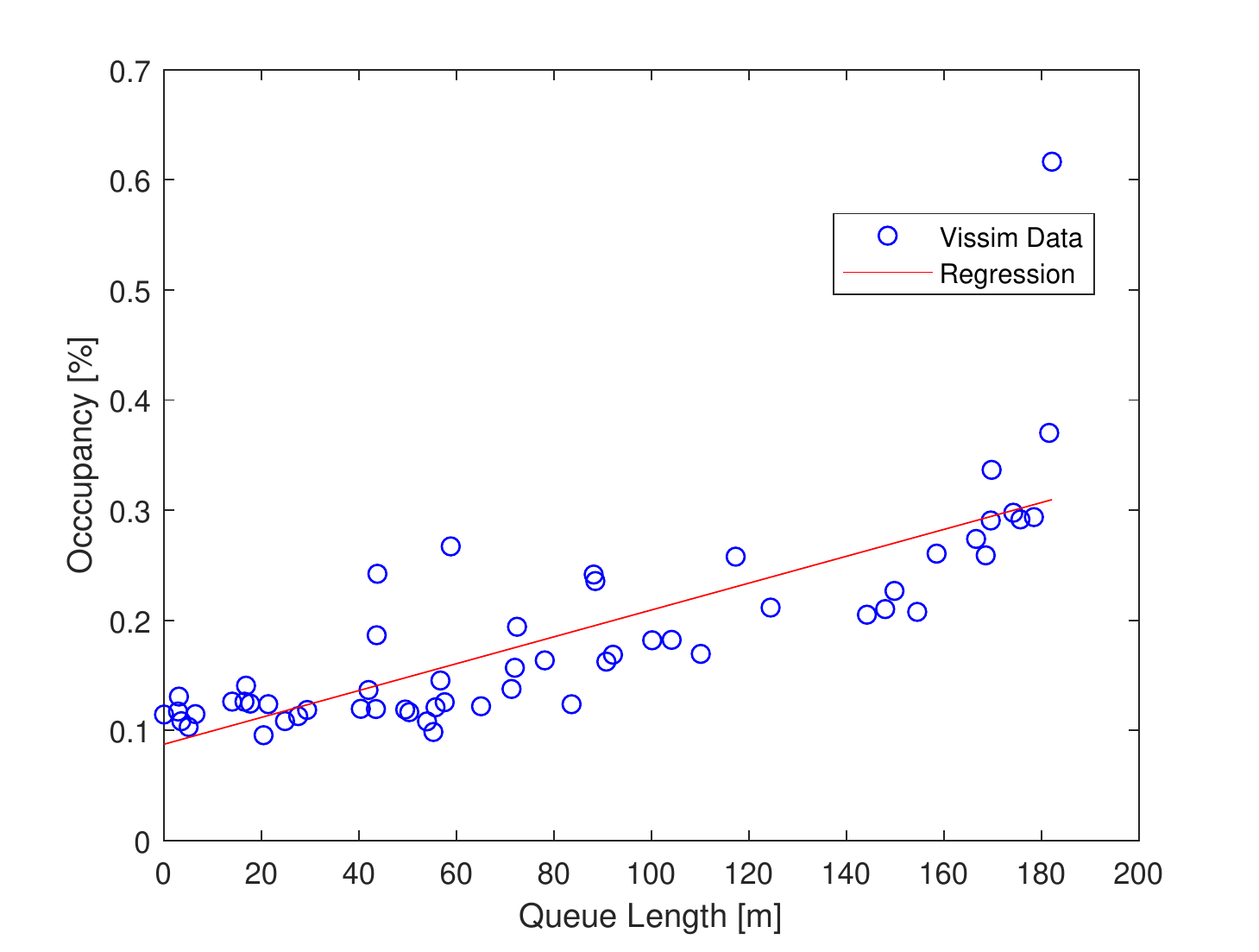}\\
(a)&(c)&(d)\\
\end{tabular}
\caption{(a) A typical smart traffic management system for an isolated four-way intersection, (b) VISSIM queue simulation, (c) Linear regression model to estimate $\lambda$, $k$ and $\beta$ using queue formation from VISSIM simulation, (d) Observables of the system with/without attack, Occupancy and Queue Length (the value of the scaling parameter $f$ is set to $0.7$ in this graph).}
\label{fig::figure_1} 
\end{figure*}

The objective of the controller is to collectively minimize the vehicle queues in each incoming lane of our intersection. Eq.~\ref{eq::ss_model} describes how queue lengths can be estimated from current queues in the system as well as the current sensor measurements
%
\begin{subequations}
\label{eq::ss_model}
\begin{align}
     Q_{t+1} =  \delta_{t}\,Q_{t} - \left[\delta_{t}\,S+\left(1-\delta_{t} \right)I_{t}\right]z_{t}+I_{t}, \\
     O_{t+1} =  k_{t}Q_{t}+\beta_{t}\,O_{t}+\lambda_{t},
\end{align}
\end{subequations}
where $\lambda$, $k$ and $\beta$ are environment-dependent parameters that describe the traffic flow within each lane and have to be estimated real-time as the traffic flow intensity changes. Here, we estimate the values of $\lambda$, $k$ and $\beta$ using our traffic flow simulator, VISSIM~\cite{fellendorf2010microscopic} by fitting a linear regression model to each arm of the intersection, as shown in Fig.~\ref{fig::figure_1}b and c.


%

The observable variables of the system such as the vehicles being dispatched at each green light and the occupancy rate measurements are given by Eq.~\ref{eq::obs_model} as
\begin{subequations}
\label{eq::obs_model}
\begin{align}
     Y_{t+1} =  \Delta Q_{t+1}+I_{t}, \\
     O_{t+1} =  O_{t+1}.
\end{align}
\end{subequations}

And finally, the control law for updating the green lights with changing occupancy rate measurements is
\begin{equation}
\label{eq::cntrl_law}
     z_{t+1} = z_{t}\, \left(1+f \times \Delta O \right), 
\end{equation}
where, the green light time, $z_{t+1}$ for the next traffic cycle depends on the current green light time $z_t$ and the change in occupancy rate measurements within the last traffic cycle. With a constant traffic flow, the green light timing settles to a value at which the occupancy rates do not change and the queues developed are constant. The parameter $f$ is a scaling factor determined experimentally.

Since turning vehicles are disregarded in our model, even though there are four incoming flows into the intersection, every flow has a complementing non-conflicting flow. That is, when the traffic lights turn green, vehicles from two lanes can access the intersection at once without causing collisions. Thus, the controller can calculate the green times as per the maximum flow intensity of the two non-conflicting flows. This will ensure that the flow with less intensity will also be served properly if the timings are decided as per the dominant flow~\cite{fhwa_handbook}. Hence, the traffic lights of the non-conflicting flows can be paired and controlled at once to improve the efficiency of the system. Thus, our traffic controller for a four-way intersection controls the timing schedule for four traffic lights or two traffic light pairs within a cycle window $T_c$. 

The equations described above can be extended for a four-way intersection traffic controller with two traffic light pairs. To ensure the robustness of our model, measurement noise ($\nu$) and process noise ($p$) are added. The complete system model equations are given below.
This traffic controller model is then implemented on MATLAB with VISSIM simulation-in-loop to verify its performance. A short video of VISSIM simulation with the adaptive traffic controller is available~\cite{video_link}.
\begin{equation}
\small
\label{eq::ss_X}
\begin{bmatrix}
Q_{1,t+1}
\\ Q_{2,t+1}
\\ O_{1,t+1}
\\ O_{2,t+1}
\end{bmatrix}=
A\times
\begin{bmatrix}
Q_{1,t+1}
\\ Q_{2,t+1}
\\ O_{1,t+1}
\\ O_{2,t+1}
\end{bmatrix}+
B\times
\begin{bmatrix}
 z_{1,t}\\
 z_{2,t}
\end{bmatrix}+
F+p_{t},
\end{equation}

\begin{equation}
\label{eq::ss_Y}
\begin{bmatrix}
Y_{1,t+1}
\\ Y_{2,t+1}
\\ O_{1,t+1}
\\ O_{2,t+1}
\end{bmatrix}=
C\times 
\begin{bmatrix}
Q_{1,t+1}
\\ Q_{2,t+1}
\\ O_{1,t+1}
\\ O_{2,t+1}
\end{bmatrix}
+
H+\nu_{t+1},
\end{equation}
where
\begin{equation*}
\small
A = 
\begin{bmatrix}
 \delta_{1,t}&  0&  0&  0 \\ 
 0&  \delta_{2,t}&  0&  0  \\ 
 k_{1,t}&  0& \beta_{1,t}& 0 \\ 
 0&  k_{2,t}&  0& \beta_{2,t} \\ 
\end{bmatrix},
\end{equation*}
\begin{equation*}
\small
B = 
\begin{bmatrix}
 -\delta S_{1,t}/dt - (1-\delta)I_{1,t}/dt& 0\\
 0& -\delta S_{2,t}/dt - (1-\delta)I_{2,t}/dt\\
 0& 0\\
 0& 0
\end{bmatrix},
\end{equation*}
\begin{equation*}
\small
F = 
\begin{bmatrix}
 I_{1,t}dt\\
 I_{2,t}dt\\
 \lambda_{1}\\
 \lambda_{2}
\end{bmatrix},
C = 
\begin{bmatrix}
0& -\alpha& 0& 0\\
\alpha& 0& 0& 0\\
0& 0& 1& 0\\
0& 0& 0& 1\\
\end{bmatrix},
H = 
-C\times
\begin{bmatrix}
 Q_{1, t} + I_{1,t}dt\\
 Q_{2, t} + I_{2,t}dt\\
 0\\
 0
\end{bmatrix},
\end{equation*}
where dt shows the simulation step time.
\begin{equation}
\resizebox{0.30\textwidth}{!}{$
\label{eq::ss_Z}
\begin{bmatrix}
Z_{1,t+1}
\\ Z_{2,t+1}
\end{bmatrix}=
\begin{bmatrix}
Z_{1,t}
\\ Z_{2,t}
\end{bmatrix}\times \left(1+f\times
\begin{bmatrix}
 \Delta O_{1}
\\ \Delta O_{2}
\end{bmatrix} \right)$} \space;
Z_{1,t+1}+Z_{2,t+1}<=T_c. 
\end{equation}

In Eq.~\ref{eq::ss_Y}, $\alpha$ represents the ratio of left turning vehicles in a given incoming traffic stream. To extend the model for turning lanes, $\alpha$ values can be found through various data collection points and traffic flow estimations. For simplicity, we set $\alpha$ to zero.

\section{Threat Model and Attack Strategy}
\label{sec::threat_model}
In a typical CPS, adversaries can attack either the analog stimulus that excites the sensor, the analog output of the sensors or the digital output of the sensor system that is fed into the control system. This paper focuses on the digital output of the inductive loop detector as a possible foothold for a replay attack.

\subsection{Attack Goal and Assumptions}
Traffic congestion can have network level effects and a large-scale impact on the safety of everyone involved. The attacker's objective is maximizing the queue length to eventually disrupt the transportation network. Hence, our attacker's goal is to decrease the efficiency of traffic lights by manipulating the traffic controller.

Since we focus on an isolated traffic intersection, this paper assumes that the traffic has not been affected by prior events such as impending queues, traffic congestion etc. Also, the vehicles arrival pattern in each lane is modeled by Poisson distribution~\cite{oppenheim1999discrete} where the length of each vehicle is the same. All vehicles are assumed to follow the rules of the traffic light. 

The attacker has the ability to view and affect the perceived occupancy rate. Once the occupancy sensors perform the measurements, the attacker is able to record the output of the sensor and sending it later at an opportune moment. The attacker may do so by attaching an additional off-the-shelf microcontroller to read the sensor values and/or an inexpensive transmitter antenna to send out the falsified data, which summarizes our replay attack. Additionally, the attacker has sufficient resources to monitor the system (intersection). They are also able to wait until the proper conditions, from their perspective, are present for the attack to take place. Critically, the attacker does not have the ability to tamper with the sensor data and can only replay the existing data. 

\subsection{Attack Strategy}
Since the control law for the traffic lights relies on the difference in occupancy rate measurements, a negative occupancy rate difference would mean reduction in traffic and would require reduction in the green light time allocation. Hence, the attacker can constantly capture the sensor data by monitoring the system and induce an attack by replaying a low occupancy measurement in place of the actual reported. By judiciously choosing the right data to send, the attacker can constantly introduce negative difference in occupancy rate measurement and disrupt the lights by eventually bringing the green light time to zero.
\begin{algorithm}
\caption{Replay attack}
\label{alg::attack}
\begin{algorithmic}[1]
\State $z$: green light duration
\State $O$: occupancy of the inductive loop detector 
\State $f$: environment dependent parameter 
\State $\mathrm{ras}$: \underline{r}eplay \underline{a}ttack \underline{s}ignal 
    \Procedure{replay attack }{$O_{t}$, $z_{t}$, $f$}
    \State $\mathrm{ras} \gets$ Inf
        \For{each $t$ in $T_{C}$ }
            \If{$O_{t}-O_{t-1} < \mathrm{ras} - O_{t-1}$}
                \State $ \mathrm{ras} \gets O_{t}$ 
                \Comment{$ \mathrm{ras}$: the smallest value of the sensor over time}
                \State $z_{t} \gets z_{t-1}  \left(1+f \left(\mathrm{ras} - O_{t-1}\right)\right)$ 
            \Else
                \State $z_{t} \gets z_{t-1}  \left(1+f \left(O_{t} - O_{t-1}\right)\right)$ 
            \EndIf
        \EndFor
    \EndProcedure
\end{algorithmic}
\end{algorithm}
In the state space model, we target the green light control law given at Eq. \ref{eq::obs_model} to craft a replay attack. This attack is performed by changing the green light equation as, 
\begin{equation}
    \label{eq::attak_model}
    z_{t+1} = z_{t}  \left(1+f \left(\mathrm{ras}-O_{t}\right) \right),
\end{equation}
where "$\mathrm{ras}$" is the \underline{r}eplay \underline{a}ttack \underline{s}ignal, which holds the smallest value of $O$ up until the current time instant, and gets updated if $O_{t+1} - O_{t} < \mathrm{ras} - O_{t}$. This is summarized in Alg.~\ref{alg::attack}. Eventually this drives the green light allocation to be less that sufficient and creates a long queue in at least one arm. 


The effect of the attack on a detector placed in one of the lanes in the system is given in Fig.~\ref{fig::figure_1}d in terms of occupancy and queue length with 700 v/h intensity as the dominant traffic flow when the attack is launched at 420 seconds. It is clear that when the system is under attack, the occupancy difference injected by the attacker makes the green light allocation smaller. Due to this, the queue length increases drastically, the occupancy sensors saturate and thus, disrupt the traffic network.     


We next evaluate the performance of our attack in presence of the state-of-the-art attack detectors.
\section{Countermeasures}
\begin{algorithm}
\caption{Threshold-based attack detection}
\label{alg::cm2}
\begin{algorithmic}[1]
\State $z$: green light duration
\State $I$: intensity of flow of vehicles into an intersection
\State $O$: occupancy of strategic loop detector 
\State $\mathrm{thrsh}$: offline value defined on the queue length in each arm
\State $\zeta$: scaling factor
    \Procedure{Define an offline threshold }{$O_{t}$, $z_{t}$}
        \State $\mathrm{thrsh} \gets -\infty$ 
        \For{each $\mathrm{t}$ in $T_{C}$ }
            \If{$\left(O_{t}-O_{t-1}\right) > \mathrm{thrsh}$}
                \State $\mathrm{thrsh} \gets \mathrm{avg}\{\mathrm{thrsh},O_{t}\} $
            \EndIf 
         \EndFor
     \EndProcedure
    \Procedure{counter measure }{$O_{t}$, $z_{t}$, $f$,$I_{t}$}
        \For{each $\mathrm{dt}$ in $T_{C}$ }
            \If{$\left(O_{t}-O_{t-1}\right) < \mathrm{thrsh}$}
                \State attack detected 
                \State $\zeta \gets \frac{\mathrm{T_c}}{\sum_{arm} I} $
                \State $z_{t} \gets z_{t-1}  \left(\zeta \times I_{t}\right)$ 
            \Else 
                \State $z_{t} \gets z_{t-1}  \left(1+f \times \mathrm{\Delta O}\right)$
                \State update offline threshold
            \EndIf 
         \EndFor
     \EndProcedure
\end{algorithmic}
\end{algorithm}

In this section, by using an approximate model for estimating the occupancy previously presented in Sec.~\ref{sec::system_model}, we show that the state-of-the-art model-based detectors fail to detect our replay attack, and then we provide countermeasures to modify such detectors.
\label{sec::countermeasures}

\begin{figure*}[t]
\centering
\subfloat[Noise power: 0.015\% of signal power]
{
\includegraphics[width=0.33\textwidth]{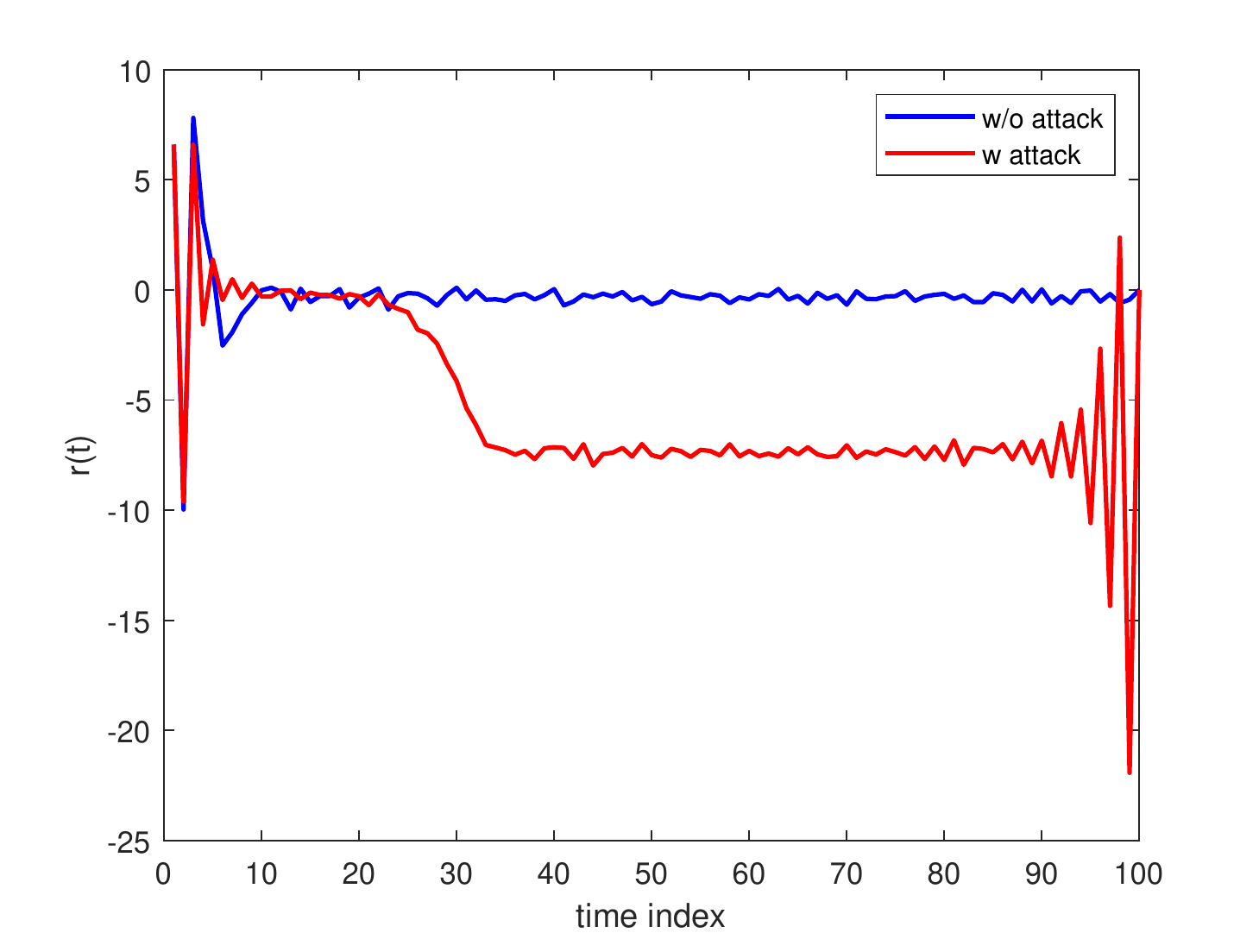}
\label{fig::1}
}
\subfloat[Noise power: 0.172\% of signal power]
{
\includegraphics[width=0.33\textwidth]{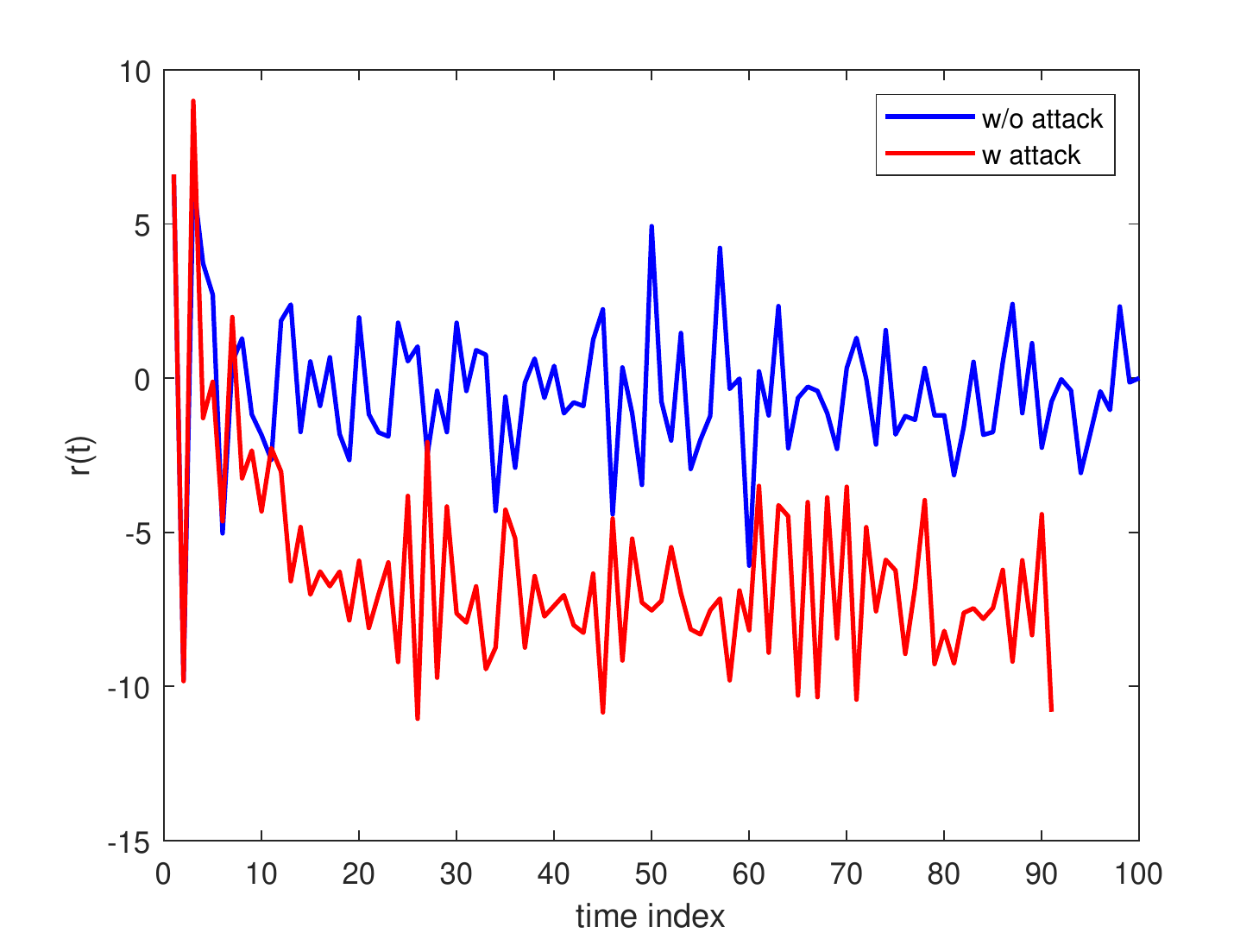}
\label{fig::2}
}
\subfloat[Noise power: 0.394\% of signal power]
{
\includegraphics[width=0.33\textwidth]{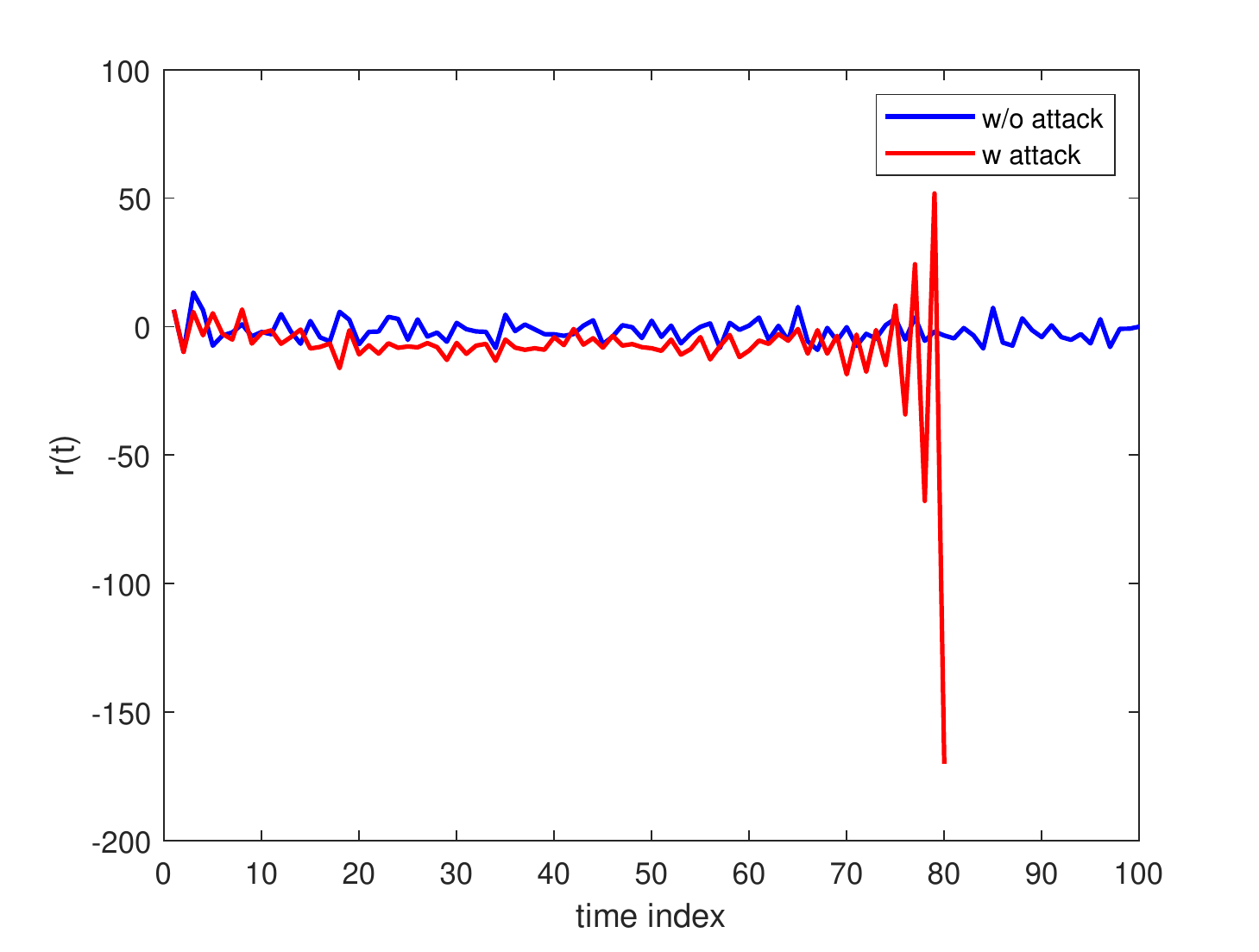}
\label{fig::3}
}
\caption{The performance of the state-of-the-art model-based detector \cite{pasqualetti2013attack} under different measurement noise power values.}
\label{fig::model-based-det}
\end{figure*}

\begin{figure}[t]
\centering
\includegraphics[width=0.8\columnwidth]{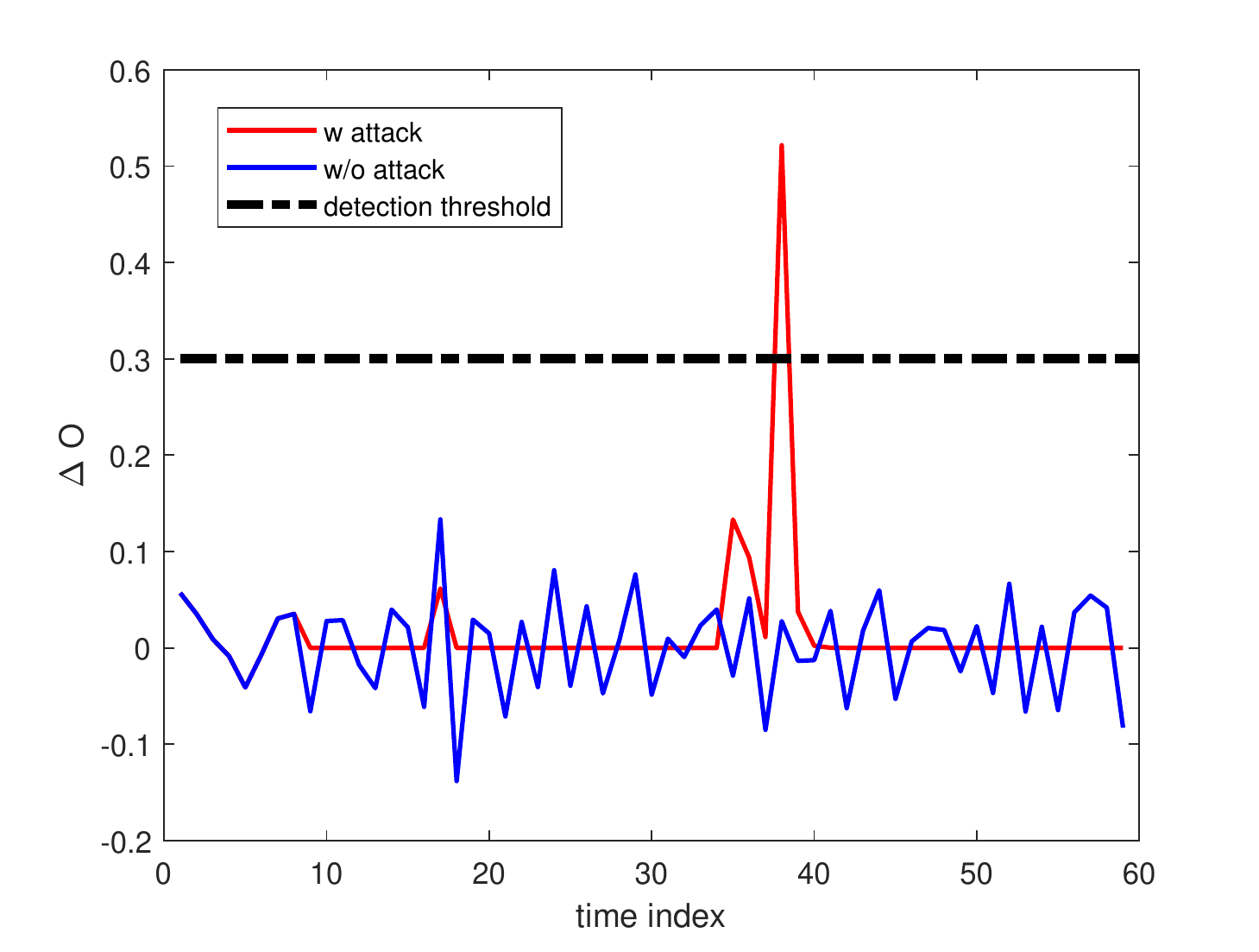}
\caption{Setting a threshold to detect an attack using the occupancy sensor values.}
\label{fig:cm2}
\end{figure}
\subsection{Model-based attack detectors}
\label{sec::cm:model}
In practice, an attack detector is based off an approximate model of the system, as it must closely represent the system without relying on having an exact knowledge of it.  The model given in Eq.~\ref{eq::appX} and Eq.~\ref{eq::appY} holds the occupancy values as the main observable in the system. The simplicity of the model, and the fact that these occupancy measurements are the main foothold for the attacker, makes the approximate model a better candidate for our analysis hereafter. The state space equations of the approximate model are
\begin{equation}
\small
\label{eq::appX}
\begin{bmatrix}
O_{1,t+1}
\\ O_{2,t+1}
\end{bmatrix}=
A' \times
\begin{bmatrix}
O_{1,t}
\\ O_{2,t}
\end{bmatrix}+ B'\left( dt \right) \times 
\begin{bmatrix}
 u_{1,t}
\\ u_{2,t}
\end{bmatrix} +
\begin{bmatrix}
 p_{1,t}
\\ p_{2,t}
\end{bmatrix},
\end{equation}

\begin{equation}
\small
\label{eq::appY}
\begin{bmatrix}
O_{1,t+1}
\\ O_{2,t+1}
\end{bmatrix}=
C' \times
\begin{bmatrix}
O_{1,t}
\\ O_{2,t}
\end{bmatrix}+ D'\left( dt \right) \times 
\begin{bmatrix}
 u_{1,t}
\\ u_{2,t}
\end{bmatrix} +
\begin{bmatrix}
 \nu_{1,t}
\\ \nu_{2,t}
\end{bmatrix},
\end{equation}
where
\begin{equation*}
A' = 
    \begin{bmatrix}
    1 &0
    \\0  &1
\end{bmatrix},
B' = 
    \begin{bmatrix}
    3 &0
    \\0  &3
    \end{bmatrix},
    C' = 
    \begin{bmatrix}
    2.5 &0
    \\0  &2.5
    \end{bmatrix},
    \\
D' = D,
\end{equation*}
and $u$ shows the input approximation as per vehicles entering and leaving and $u_{t+1}$ is given as
\begin{equation}
  u_{t+1} = i \times dt - S \times z_{t+1}.
\end{equation}

Next, we explore the existing methods for detecting our replay attack on the approximate model. Model-based attack detectors are extensively covered in the state-of-the-art as strong candidates for detection of false-data injection in control systems \cite{pasqualetti2013attack}. Here we demonstrate shortcomings of the model-based approach by Pasqualetti et al.~\cite{pasqualetti2013attack} in detecting our replay attack, and then provide our threshold-based countermeasure which successfully detects malicious activities when our attack is in progress. Our results are then validated through VISSIM.  

Model-based attack (anomaly) detectors for CPS are those that make an estimate of how the system should be observed to evolve and compare it with the measured value of the state (how the system actually evolves), shown by $r\left(t\right)$ at Eq.~\ref{eq::monitor}. In absence of an attacker $r\left(t\right)$ should be close to zero. Pasqualetti et al~\cite{pasqualetti2013attack} propose a monitor $\omega\left(t\right)$ for this purpose which is defined as
\begin{equation}
\label{eq::monitor}
    \begin{aligned}
        E\, \dot \omega \left(t\right) &= \left(A'+G\,C'\right) \omega\left(t\right) - G\,y\left(t\right), \\
        r\left(t\right) &= C'\, \omega\left(t\right) - y\left(t\right),
    \end{aligned}
\end{equation}
where $E=1$ in our approximate model, $\omega\left(0\right) = x\left(0\right)$, and $G$ is the output injection matrix which should be selected so that the pair $\left(E,A'+G\,C'\right)$ is regular and Hurwitz~\cite{hurwitz1898ueber}.  

Fig.~\ref{fig::model-based-det} illustrates the robustness of the Pasqualetti's detector to the measurement noise power when our attack is taking place. With low measurement and process noise, an abrupt change will appear in the error calculations which makes it possible to perform anomaly based attack detection shown in Fig.~\ref{fig::1}. As the measurement noise power increases while the estimation noise is kept consistent, the performance of Pasqualetti's detector degrades (Fig.~\ref{fig::2}) and eventually breaks (Fig.~\ref{fig::3}). This shows that our replay attack is not detectable by the model-based detector of the state-of-the-art if noise power exceeds $0.394\%$ of the original signal power.

\subsection{Threshold-based attack detectors}
\label{sec::cm:thrsh}
Having shown the shortcomings of the model-based attack detectors to detect our replay attack, we now introduce a countermeasure in this section.
The conventional and most common countermeasure against false-data injection attacks on control systems is Kalman Filter (KF) estimation of the states of the system followed by a simple comparison \cite{dunik2006state,mo2010false}. Even though KF has high accuracy of estimation over the other available filters, it is designed on the basis of the not necessarily correct assumption that the parameters are Gaussian distributed in a linear system. Other filters such as particle filtering \cite{ward2003particle} overcome this shortcoming by addressing the general non-Gaussian non-linear systems at the cost of higher complexity. To avoid the complexity-performance trade-off, a general solution to this problem could be using the timestamps of the inductive loop detector output. However, this method depends on another sensor output and there is no guarantee that it is secure. Also, a sophisticated attacker can forge time-stamps by crafting a more complicated attack than a replay attack. 

We address these shortcomings by proposing a threshold based countermeasure described in detail in Alg.~\ref{alg::cm2}. The threshold is defined in an offline training process to distinguish between benign and malicious activities by looking into the changes in the occupancy sensors readings in two consecutive time instants (given by $\Delta O$). In this case, we adjust the threshold in a way that it detects the abrupt change generated in the reported $\Delta O$ when there are ongoing malicious activities. Anytime the change in $\Delta O$ goes above this threshold an attack is said to be detected. This threshold changes overtime, and needs to be updated using secure data. Note that upon detection of an attack the green light values get updated by a  new control law
\begin{equation}
   \label{eq::new_glight2}
    z_{t}=z_{t-1}  \left(\zeta \times I_{t}\right),
\end{equation}
where $\zeta$ is a normalizing factor given by $\zeta=\frac{\mathrm{T_c}}{\sum_{arm} I}$, such that the green time is proportionally distributed as per the incoming traffic flow rate, while assuming the $I_t$ measurement is secure.

Fig. \ref{fig:cm2} shows the change of occupancy over time for one arm with/without attack, where the threshold is defined to create separability between the spike in the graph caused by the attack versus the case with no attack. 
\subsection{Time Complexity}
The threshold-based detection mechanism is provided in Alg.~\ref{alg::cm2}, where the training of threshold is performed offline and hence does not add to the time complexity of the approach. The detection mechanism however, includes a comparison operation which results in time complexity of $O\left(n\right)$.

\section{Conclusion}
\label{sec::conc}

In this paper, we first introduce a replay attack on the inductive loop sensors which are used for traffic control in an isolated four-way intersection. More specifically, the occupancy sensor measurements are used as a feedback to adjust the green light timings for the traffic lights. By introducing a replay attack, a type of false-data injection attack, our attacker minimizes the efficiency of the traffic controller by affecting the green light allocation and thereby maximizing the queue lengths in at least one of the lanes of the intersection. By showing that the existing model-based attack detectors fail in detecting our attack, we put forth an alternative threshold-based detection algorithm. This detector observes the occupancy sensor data over time and raises a flag if the changes in the occupancy sensor measurements are above an offline defined threshold. Assessment of the current threshold-based detector under a smarter attacker capable of evasion attacks is the future direction of our work. We are also interested in extending this study to analyse the feasibility of our approach for broader ranges of traffic models that rely on occupancy sensor measurements.

\section{Acknowledgement}
\label{ack}
This work was supported in part by the U.S. National Science Foundation under grant number 1658225.

\bibliographystyle{ACM-Reference-Format}
\bibliography{bibliography}


\end{document}